\documentclass[twocolumn,showpacs,preprintnumbers,amsmath,amssymb]{revtex4}

\usepackage{graphicx}
\usepackage{dcolumn}
\usepackage{bm}

\begin{document}

\preprint{APS/123-QED}

\title{Dust-Induced Destabilization of Glacial Climates}
\author{Brian F. Farrell}
 \email{farrell@seas.harvard.edu}
\affiliation{%
Harvard University\\
}%

\author{Dorian S. Abbot}
\email{abbot@fas.harvard.edu}
\affiliation{
Harvard University\\
}%

\date{\today}

\begin{abstract}
The climate record  preserved in polar glaciers, mountain glaciers, and widespread cave deposits shows repeated occurrence of abrupt global transitions between cold/dry stadial and warm/wet interstadial states during glacial periods.   These abrupt transitions occur on  millennial time scale and in the absence of any known global-scale forcing. Here a theory is advanced implicating a feedback between atmospheric dust and the hydrological cycle in producing these  abrupt transitions.  Calculations are performed using a radiative-convective model that includes the interaction of aerosols with radiation to reveal the mechanism of the dust/precipitation interaction feedback process and a Langevin equation  is used to model glacial climate destabilization by this mechanism. This theory explains the observed bimodal, stochastic, and abrupt nature of the transitions as well as their intrinsic connection with the hydrological cycle. 
\end{abstract}

\maketitle

\section{Introduction}
A central problem in climate science is understanding the mechanism producing abrupt climate change  such as   the glacial/interglacial  transitions and the Dansgaard-Oeschger interstadial transitions  during the glacial. The rapidity with which the glacial climate state changes argues for an underlying  nonlinear mechanism taking the form of a switch or trigger \citep{Clement:2008,Ditlevsen:2000}.   Climate records from ice and deep sea cores establish a relation between Milankovitch  cycles and glacial climate variability but not a deterministic causal connection \citep{Wunsch:2004}.  Additional mechanisms must be involved in producing climate variability and the  millennial time scale climate variation recorded in Greenland ice \citep{Alley:1993} and Chinese cave deposits \citep{Wang:2001} shows that these rapid climate variations do not depend on orbital parameter, ice albedo or CO$_2$ variation, which have longer time scales.  In addition to the the decadal and shorter time scale of these transitions, particularly telling is  their hemispheric to global spatial scale  and the intimate association of these events with the hydrological cycle implied by polar records of  ice accumulation \citep{Steffensen:2008,Alley:1993}, dust and aerosol loading \citep{Lambert:2008,Fuhrer:1999}, oxygen isotope  and methane \citep{Barnola:2008,Chappellaz:1993} which are corroborated by  midlatitude glacier records \citep{Thompson:1989} and widespread cave oxygen isotope records \citep{Wang:2001,Fleitmann:2003,Yuan:2004,Dykoski:2005}.

Rapid increase in atmospheric methane during abrupt warming events implies tropical precipitation changes on time scales as short as three to six years  coincident with equally rapid variation in the midlatitude Asian monsoon precipitation intensity  \citep{Bigler:2007} and Greenland ice accumulation rates   \citep{Mayewski:1993}.  It was remarked early in the study of these events that atmospheric dust varies by at least an order of magnitude on the same decadal time scale  \citep{Fuhrer:1999}.  It is also well-established that a great deal of dust was in flux during glacial periods producing widely distributed loess deposits  \citep{Sun:2006}  and that during glacial periods the Peistocene megafauna roamed vast areas of a dusty Mammoth steppe in what is now wet permafrost tundra  \citep{DaleGuthrie:2001}.   Large increases in atmospheric dust load observed during these cycles  \citep{Thompson:1989,Lambert:2008}
is usually explained as  a consequence of the reduced hydrological cycle  \citep{Yung:1996} and the observed  decrease in dust coincident with warming  is consistent with return of pluvial conditions to source regions and increase in atmospheric dust wash out in association with a stronger hydrological cycle.

Clues to the mechanism of abrupt climate change are provided by these records.  The bimodality, abrupt onset and memoryless nature  of the change implies random occurrence of a switch-like transition between states.  The hemispheric scale, extreme rapidity and concurrent change in the hydrological cycle at both low and mid latitudes implies this switch-like rearrangement in the state of the atmosphere  is related to precipitation  on a hemispheric scale \citep{Rohling:2003}. Moreover, this rearrangement must involve strong enough feedbacks to produce and maintain  two stable states, one dry/cold and one wet/warm.  

\section{The Mechanism of Dust Destabilization}
A suggestive analogue of abrupt transitions seen in Earth's glacial climate  has been observed on Mars where transition between a clear and a dusty state occurs with a bias toward the dusty state transition during Martian northern summer  \citep{Fernandez:1997}.  In fact, the analogy with Martian dust storms is apt as the Earth's climate also has a crucial and vulnerable interaction among atmospheric dynamics, dust, and radiation; with the additional involvement of the hydrological cycle in the case of the Earth.  The interaction between radiation and the hydrological cycle drives the monsoons and the instability of this interaction is notorious in limiting monsoon prediction. With these observations and concepts in view we next examine more closely the influence of dust on glacial climate variability. 

Calculations with a radiative-convective model show that both atmospheric reflection and absorption by dust reduce radiation reaching the surface (Fig.~1b), which exerts a strong influence on global precipitation rate (Fig.~1a) essentially by reducing the energy available to evaporate water.  A more subtle influence of radiation absorption on precipitation becomes important when the dust is increased to 5-20 times modern levels (State 2 in Fig.~1,2). Absorption of radiation by dust begins to replace vertical transport of latent energy by convection  in maintaining the marginal convective stability of the troposphere (Fig.~2c). This cuts off mid-tropospheric dust from wet deposition, eliminating the primary process by which dust is removed from the atmosphere. For example, when dust is increased to ten times modern levels, deep convective precipitation production is reduced by a factor of 2.6 at 607 mb (Fig.~2c), the altitude of maximum dust mixing ratio (Fig.~2a). This drastic reduction in the dust sink means that dust input would only need to be increased by a factor of 2-8 for the midtropospheric dust concentrations to reach 5-20 times modern levels so that this state could persist, well within the range of estimated dust source increase during cold Pleistocene periods  \citep{Thompson:1989,Lambert:2008,Fuhrer:1999,Winckler:2008}. If the dust is further increased to $\sim$30 times modern, an inversion develops near the altitude of maximum dust concentration (State 3, Fig.~2b, 1a). In this state deep convection shuts off completely (Fig.~2c) and total precipitation is reduced by more than an order of magnitude (Fig.~1a). 

The state in which the role of convective latent heat transport and precipitation in maintaining the marginal convective stability of the troposphere is partially assumed by absorption of radiation by dust (State 2) is a model for the large-scale behavior of the atmosphere during cold, dry, glacial climates.
Switch-like transitions between the cold, dry, dusty glacial climate (State 2) and the warm, wet, less dusty interglacial climate (State 1) occur  when atmospheric dust load exceeds a threshold resulting in switch-like behavior as indicated in Figs.~1 and 2.  Crossing this threshold would depend on a sequence of weather events that result in sufficiently high atmospheric dust load.  Dust load is a strong function of surface drying which in turn is related  to the particular sequence of global weather events that happens to  occur.   Even with identical boundary conditions individual realizations of global weather events result in large integrated dispersion in continental scale drying as seen among realizations in ensemble AGCM simulations  \citep{Schubert:2004}.  In addition dust lofting is a strong nonlinear function of surface wind speed  \citep{Andersen:1998} which also varies among ensemble members in simulations both with and without boundary condition influences.  The effect of changes in atmospheric boundary conditions, such as are associated with  ENSO and the  PDO in a coupled AOGCM,  is to produce  an additional stochastic variation in regional weather patterns that serves to enhance the variance of the stochastic process that  triggers the dust state transition.  Explicit simulation of the dust state transition is not possible in the present generation of climate models since they do not solve for dust as a prognostic variable and since dust lofting depends crucially on meso and micro scale processes  \citep{Schepanski:2009} that they do not resolve. Therefore the calculations presented here, while indicative, can not establish precisely a sequence of events that would cause such a transition, although we can say that the occurrence would be quite random.

The switch-like nature of the dust transition and the highly nonlinear relation between wind speed and dust lofting imply in the free running climate system  random occurrences of widespread excursions in  dust loading triggering associated abrupt transitions even in the absence of external influences. This is consistent with a memory-less  Poisson process characterized only by a time scale for transitions as is seen in the glacial climate record.  Consistent with these statistical properties, the glacial climate can be phenomenologically modeled using a Langevin equation with a pseudo-potential having two minima representing the two stable states separated by a potential barrier and forced by white noise \citep{Ditlevsen:2000}.  With suitable parameter choice a stochastic model of this kind produces a time series of dust concentration in good agreement with the ice core record.  The contribution of our dust-induced transition theory to this Langevin  model for abrupt transition is to make a physical identification of the two phenomenological states  as being the warm/wet state and cold/dry state and to identify the pseudo-potential as  corresponding to the switch-like dust-induced instability between these states.  
As an illustrative example, using a state variable of $x=log_{10}(dust)$, we choose the interstadial state to correspond to the present dust level ($x=0$) and a typical stadial dust level of 10 times the present dust level ($x=1$).  We then choose a potential ($U$) corresponding to these equilibria, with a switch between them at  $x = 0.5$ (Fig. 3a). The corresponding Langevin equation is
$$
\dot x=-\frac{dU}{dx} + \sigma \xi,
$$
with $\sigma = .1$ and $\xi$ Gaussian distributed white noise with zero mean and unit variance. This equation produces a time series (Fig. 3c) in which the system abruptly switches between states with exponentially distributed waiting time indicative of a stochastic Poissson process, consistent with the glacial records.  We have chosen to put the glacial equilibrium at 10 times present dust in agreement with observations of dust, temperature  and precipitation
in glacial climates.  We believe the far dustier and colder equilibrium indicated by the relationship between dust and precipitation in Fig. 1a is not accessible because dust levels can not increase further once all potential dust source regions have been fully activated.  This observation suggests that the progressive increase in the severity of stadials during the Pleistocene culminating in the particularly severe LGM interval can be related to increased availability and activation of dust source regions.  Over long time scales the distribution and activation of dust source regions is clearly related to seasonal distribution of insolation and to variations in orography.

We have thus far concentrated on the mechanism underlying the stochastic component of climate variability. There is also a substantial component of variability in phase with orbital forcing.  Weak, but deterministic, external influences such as orbital-parameter-induced variation in seasonal insolation can produce clear signals in stochastically forced  systems such as the dust-destabilized glacial climate  by the process of stochastic resonance  \citep{Gammaitoni:1998}.  This provides an explanation for the appearance of orbital periodicities in the climate record even though the  effects of these insolation changes are insufficient to directly drive the climate system.  In particular  the precessional cycle, which strongly controls the Monsoon, influences dust input from large regions of the planet and would be expected to show prominently in the record of dust induced climate variation.  The highly nonlinear nature of the dust feedback is consistent with the precessional cycle forcing at $\sim$20 kyr producing both the early Pleistocene $\sim$40kyr and late Pleistocene $\sim$100 kyr glacial cycles if these are seen as nonlinear responses to cyclic forcing, which characteristically results in power appearing at some integer fraction of the forcing signal frequency  \citep{Tziperman06}. The Himalayan uplift over the Neogene would have slowly isolated the interior of Asia from the Monsoon,  eventually drying Asia enough that precession-induced aridity variations  could interact with integrated stochastic weather variation to generate sufficient dust to initiate glacial cycles.  The long-standing puzzle of the correlation between global glaciations and Northern Hemisphere insolation variations can be explained by the influence of the dust-induced precipitation decrease in the tropics which would effectively transmit the influence across the equator. 



\section{Discussion}
Understanding  past climate change presents  a fundamental  theoretical challenge, and predicting future climate change  is important for  society.  In pursuit of these goals particularly important issues are establishing the physical mechanisms controlling  the climate system and placing bounds on the rapidity with which climate change can occur.   Evidence  preserved in  ice cores and widespread tropical and mid-latitude ice and cave deposits reveal that glacial climates were characterized by  abrupt stochastic  bimodal transitions.  These observations  constrain mechanisms for explaining  glacial climate variability.  Abrupt transitions can be produced by  thermohaline  \citep{Weaver:1991}, sea ice  \citep{Gildor:2003}, and storm track  \citep{Farrell:2003} switches, but the effects of these transitions are  local  and fail to predict the observed abrupt and  global involvement of the hydrological cycle.   In this work we described a theory implicating the precipitation/dust feedback destabilization of the hydrological cycle in explaining  rapid climate transitions.


%
 \section*{Appendix: Radiative-Convective Model}
In order to model the mechanism of dust-induced destabilization of glacial climates we use
NCAR's single column atmospheric model (SCAM). This model contains all  aerosol, cloud, convection, and radiation representations of CAM, NCAR's atmospheric general circulation model. We couple the model atmosphere to a mixed layer ocean with a depth of 50 m, so that the equilibrated model conserves energy, which is important for  determining precipitation. We apply an ocean heat flux divergence of -70 W m$^{-2}$ to this slab ocean to represent heat transport to higher latitudes.  We apply dust to the model by multiplying the tropical-average of the seasonally-varying standard CAM aerosol dust climatology by a ``dust factor" at each model level for each of the CAM dust size bins, which represent dust particles with diameter 0.1-1.0 $\mu$m, 1.0-2.5 $\mu$m, 2.5-5.0 $\mu$m, and 5.0-10.0 $\mu$m. Each dust bin has characteristic optical properties. We use a sea salt aerosol profile taken from the tropical Pacific and set all other aerosols to zero. We apply a surface wind speed of 8 m s$^{-1}$ to the model. We find similar results with surface wind speeds of 4 and 12 m s$^{-1}$. We average model results over 20 years of converged solutions. We use a time step of 1200 s; our results are very similar when we reduce the time step to 600 s.

\begin{acknowledgments}
This work was partially supported  by NSF ATM-0736022 and NSF ATM-0754332.
\end{acknowledgments}

\clearpage
\begin{figure}
\begin{center}
 \noindent\includegraphics[width=20pc]{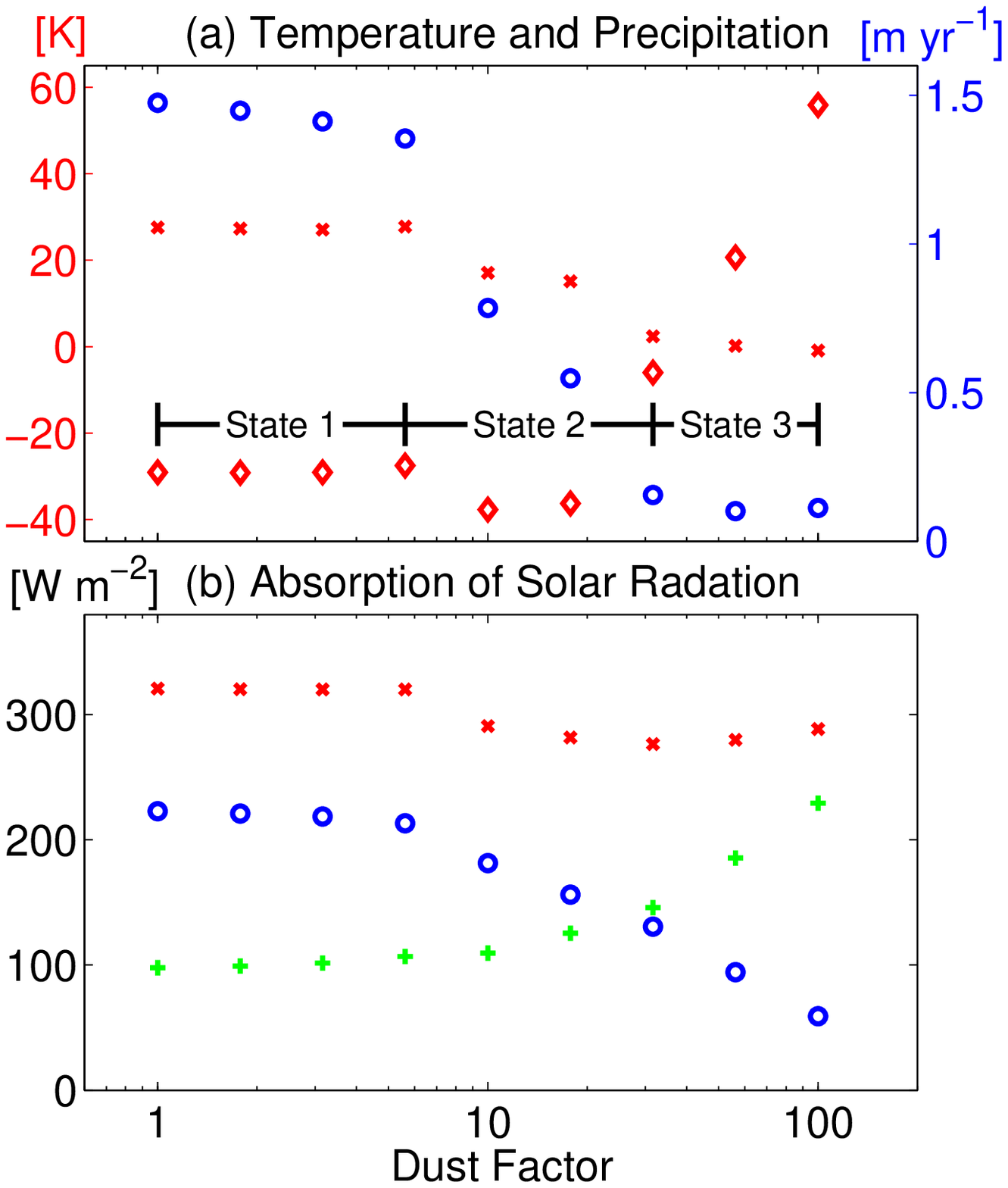} 
\end{center}
\caption{\textbf{Regimes in tropical stability and precipitation determined by atmospheric dust level.}  
(a) The precipitation (blue circles), tropical surface temperature (red x's), and the difference between mid-tropospheric (515 mb) and surface potential temperature (red diamonds) as a function of the factor by which current tropical dust levels are increased (dust factor). (b) The absorption of solar radiation by the surface (blue circles), the atmosphere (green +'s), and the sum of the surface and atmosphere (red x's) as a function of the dust factor, the factor by which modern dust levels are multiplied. In State 1 convection maintains the marginal convective stability of the troposphere. In State 2 absorption of solar energy by midtropospheric dust partially replaces convection in maintaining marginal convective stability. In State 3 solar absorption by dust causes an inversion.}
\end{figure}


\begin{figure}
\noindent\includegraphics[width=40pc]{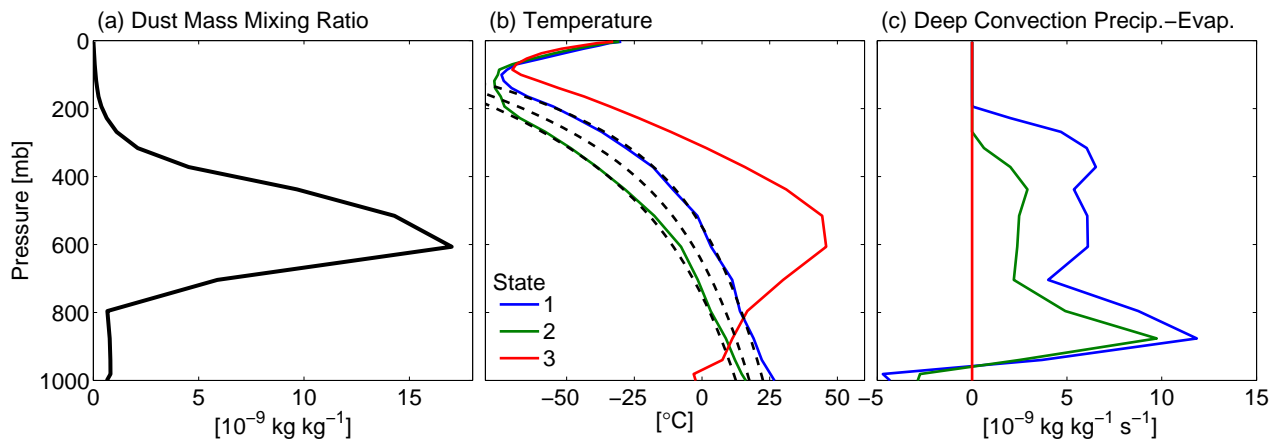} 
\caption{\textbf{Atmospheric stability and deep convection in each of the dust-mediated states.}  (a) Vertical profile of the annual-mean tropical dust mixing ratio in the current climate (dust factor of one). (b) Vertical profiles of atmospheric temperature for different dust factors. Moist adiabats with surface temperatures of 12.5$^\circ$, 27.5$^\circ$, and 22.5$^\circ$ are plotted as dashed black lines. (c) Net precipitation production (precipitation minus evaporation) by deep convection. A dust factor of 1 is used to demonstrate State 1, 10 for State 2, and 100 for State 3.}
\end{figure}


\begin{figure}
\noindent\includegraphics[width=40pc]{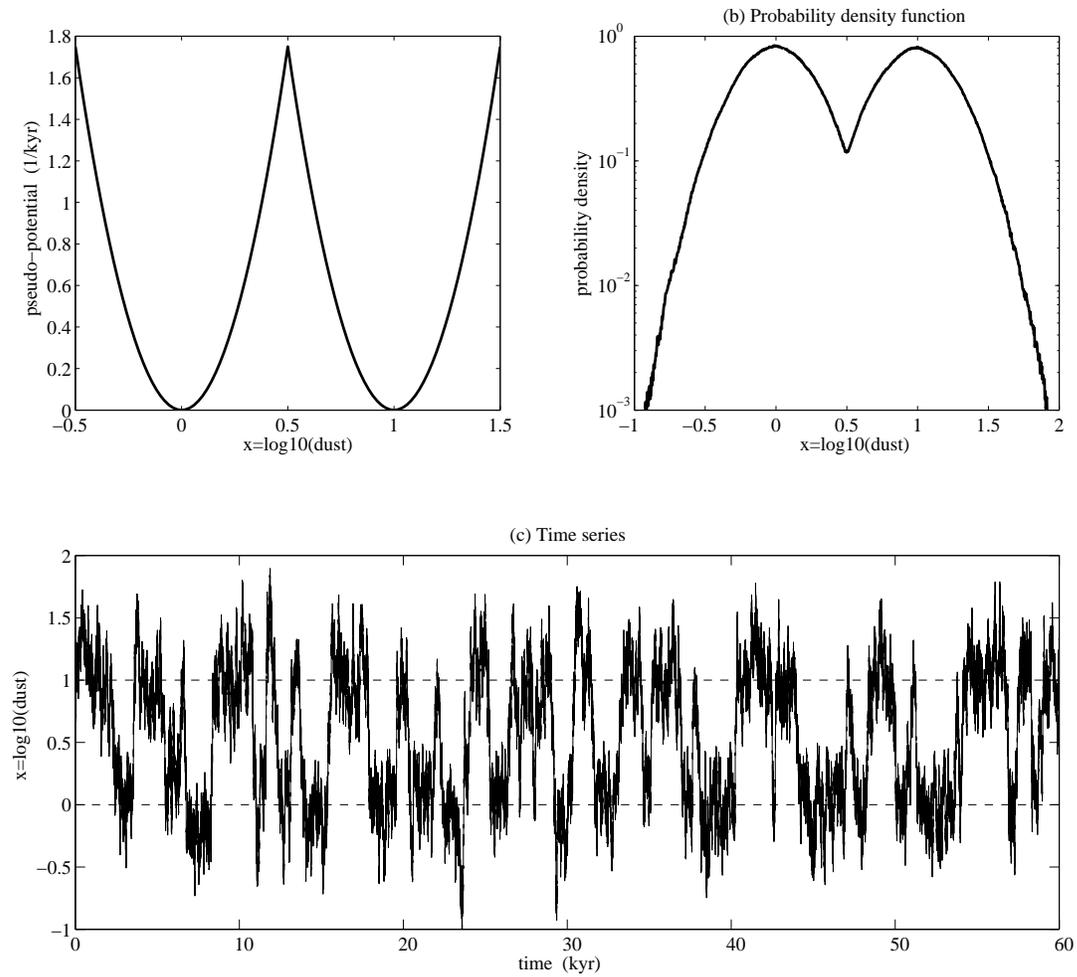} 
\caption{\textbf{Langevin Model of Dust Transitions.}  (a) Pseudo-potential corresponding to the stadial and interstadial equilibria and to the switch between them at $x = 0.5$, where $x=log_{10}(dust)$. (b)  Probability of occurrence of states.  (c) Time series of transitions between states with waiting time between transitions of 1600 years.}
\end{figure}

\end{document}